\documentclass[twoside,twocolumn,reqno]{mbook}
\usepackage{epsfig,array,amscd}
\usepackage[psamsfonts]{amssymb}
\def\bf{\bfseries}
\begin{document}

\setcounter{page}{1}

\frenchspacing

\sloppy

\markboth{\small Stoyan Pisov}{\small Short title} 

\thispagestyle{myheadings} 

\twocolumn[  

\mbox{} 

\vspace*{2\baselineskip} 

\begin{center} 

{\bf Optimization of molecular clusters configurations using a Genetic 
Algorithm}

\vspace*{2\baselineskip}

Stoyan Pisov and Ana Proykova

\bigskip

\begin{minipage}{12cm}
\small
Department of Atomic  Physics, 
Faculty of Physics, University of Sofia,  
Sofia-1126, 5 J. Bourchier Blvd.

\vspace*{3\baselineskip}

\small\noindent
{\bf Abstract}.
We present a genetic algorithm developed (GA) to optimize molecular $AF_6$ 
cluster configurations with respect to their energy. The method is based 
on the Darvin's evolutionary theory: structures with lowest energies 
survive in a system of fixed number of clusters. Two existing structures 
from a given population are combined in a special way to produce a new 
structure (child) which is kept if its energy is lower than the highest 
energy in the ensemble. To keep the population constant we reject the 
structure with the highest energy. This algorithm gives a better result 
than the optimization techniques used previously. Using the GA we have 
found a new structure corresponding to the (seemingly) global minimum. The 
most important result is that the new structure is detected only if the 
molecular cluster contains more than a critical number of molecules.
\end{minipage}
\end{center}
]
\section{Introduction}
\label{intro}
\begin{figure}[htb]
\epsfig{file=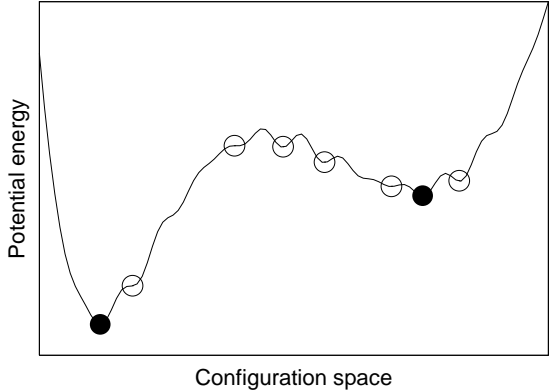, height=55mm}
\caption[]{Schematic representation  of Potential Energy Surface (PES).}
\label{pot-minima}
\end{figure}

Optimized structures give a detailed information about symmetry, phase 
transitions and we also can use them to calculate density of states of the 
studied substance. Every optimization algorithm tries to find the configuration 
with the lowest energy. Usually it minimizes the potential energy 
performing consecutive steps from one to another configuration by 
inspecting the local minima on the potential energy surface (PES). 
Molecular clusters made of octahedral molecules ($AF_6, A=S,Se,Te,U$) have 
rugged potential energy surface \cite{stoyan-01}. Attempts to use 
simulating annealing \cite{Biswas} to find the global energy minimum in 
the systems often fail due to high-energy barriers (Fig.7 in 
\cite{stoyan-heronpress}), which trap the simulated system in one of the 
innumerable metastable configurations. A cartoon can be seen in Fig. 
\ref{pot-minima}. On such a surface, techniques like simulated annealing, 
quenching \cite{Biswas}, or the conjugate gradient method find the local 
"global" (glocal) minimum  , which might lie higher than the true global 
minimum if it is localed in another basin \cite{Kunz-93}. Hence we need an 
algorithm, which would permit "jumps" from one basin to another and sample 
properly the phase space. Various techniques of global optimization have 
been proposed: basin hopping \cite{Doye-99}, genetic algorithm 
\cite{Deaven}, adiabatic switching \cite{Rama-02}.

In the present work we further elaborate GA originally 
developed of Deaven and Ho for atomic systems in order to make it 
applicable to molecular clusters. We implement the new algorithm for 
configuration optimization of $AF_6$ clusters, simulated with 
molecular dynamics \cite{stoyan-01}. A new structure has been found that 
has never been seen in our previous investigations \cite{Ani-99}, 
\cite{stoyan-01}.

\section{A genetic algorithm for molecular clusters}
\label{algorithm}
In this section we describe our genetic algorithm in detail. Each 
molecule is defined with a pair of coordinates $\{\mbox{{\bf 
x}}, \mbox{{\bf q}}\} = \vec{X}$, $\mbox{{\bf x}}=\{x,y,z\}$ is the 
Cartesian coordinate of molecular center of mass, $\mbox{{\bf q}}=\{q_0, 
q_1, q_2, q_3\}$ is molecular orientation in quaternion representation. We 
denote a cluster configuration with $N$ molecules with:
\[
\mathfrak{G} = \{\vec{X}_1, \vec{X}_2,..., \vec{X}_N\}
\]
The genetic algorithm uses a population of {\it n} structures 
$\{\mathfrak{G}\}$, {\it n} is kept constant during the optimization run. 
We define a mapping operator $P: P(\mathfrak{G, G'})\rightarrow 
\mathfrak{G''}$, which performs the following action upon two parent 
geometries $\mathfrak{G}$ and $\mathfrak{G'}$ to produce a child 
$\mathfrak{G''}$. First we select parents from the population using the 
distribution Eqn.(\ref{prob}). Second, we choose planes that account for 
the parent clusters packing symmetry and pass through the center of mass 
of the parents. Then we cut the clusters in the chosen planes. We would 
like to underline that the choice of cutting planes is crucial for 
the proper work of the algorithm. In other words, it is very important to 
find out the packing symmetry of all clusters in the populations.

In the case of solid molecular clusters at a low temperature, the centers  
of the molecules hardly move but their orientations do. Hence, in the 
present version of Genetic Algorithm, we match centers of the parent 
clusters before searching for a suitable plane. After cutting the parents, 
we assemble the child $\mathfrak{G''}$ from the molecules of 
$\mathfrak{G}$ which lie above the plane and the molecules of 
$\mathfrak{G'}$ which lie below the plane. If the child generated in this 
manner does not contain the correct number of molecules, we translate the 
plane until the child $\mathfrak{G''}$ contains the correct 
number of molecules. Relaxation to the nearest local minimum is performed 
with a conjugate gradient minimization \cite{NR,stoyan-heronpress}.

We preferentially select parents with a lower energy from 
$\{\mathfrak{G}\}$. 
The probability $p(\mathfrak{G})$ of an individual candidate 
$\mathfrak{G}$ to be selecting for mating is given by the Boltzmann 
distribution.
\begin{equation}
\label{prob}
p(\mathfrak{G}) \propto \exp[-E(\mathfrak{G})/K_bT_m]
\end{equation}
\noindent where $E(\mathfrak{G})$ is the energy of the candidate 
$\mathfrak{G}$, $K_b$ is the Boltzmann constant and $T_m$ is the mating 
"temperature", chosen to be roughly equal to the range of energies in 
$\{\mathfrak{G}\}$. For a better performance we can apply mutations to 
some members ($\mu$) of the population. The mutation operator is defined 
as $M : M(\mathfrak{G})\rightarrow \mathfrak{G'}$ which performs two 
random actions with the same probability. First, $M$ moves the coordinates 
of mass centers in a random direction with a random step. Second, $M$ 
rotates the chosen molecule at a random angle. Such mutation can be 
applied to some molecules in a cluster $\mathfrak{G}$ or to all of them.

We create subsequent generations as follows. Parents are continuously 
chosen from $\{\mathfrak{G}\}$ with a probability given by Eqn.(\ref{prob}) 
and mated using the mating procedure described above. The fraction $\mu$ 
of the children generated in this way are mutated; $\mu = 0$ means no 
mutation occurs. The (possibly mutated) child is relaxed to the nearest 
local minimum and replaced with a configuration with a higher energy in 
population $\{\mathfrak{G}\}$ if its energy is lower than the higher 
energy. This algorithm requires a great number of members ({\it n}) in the 
population in order to prevent a rapid convergence to a set of identical 
candidates.

\section{Results for $TeF_6$ clusters}
To illustrate the method, we used configurations for $TeF_6$ clusters 
obtained in MD simulations described elsewhere \cite{stoyan-01}.

\begin{figure}[htb]
\mbox{\epsfig{file=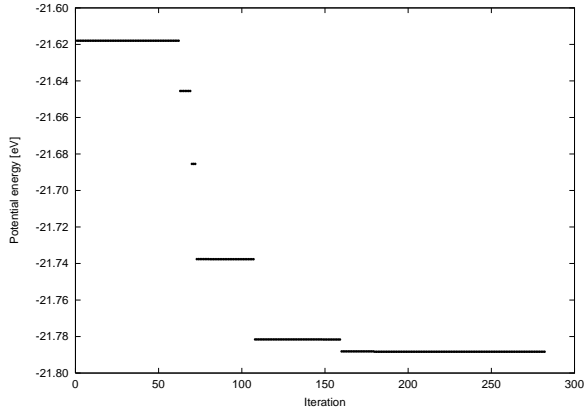, height=55mm}}
\caption[]{A population of {\it n} = 100 clusters each containing 89 
$TeF_6$ molecules finds the configuration with the lowest energy about 160 
iteration steps.}
\label{energy-run}
\end{figure}

Fig. \ref{energy-run} shows the minimum energy found in an optimization of 
n=100 configurations (starting population) of 89-molecule $TeF_6$ 
clusters. During the first 100 iterations, the algorithm effectively and 
rapidly creates better children. Then the process is slowed after the 
$100^{th}$ step. An important comment is that we start with a population 
of structures already optimized with conjugate gradient or simulated 
annealing methods. We underline that both techniques give the same 
optimized structures.

\begin{figure}[htb]
\epsfig{file=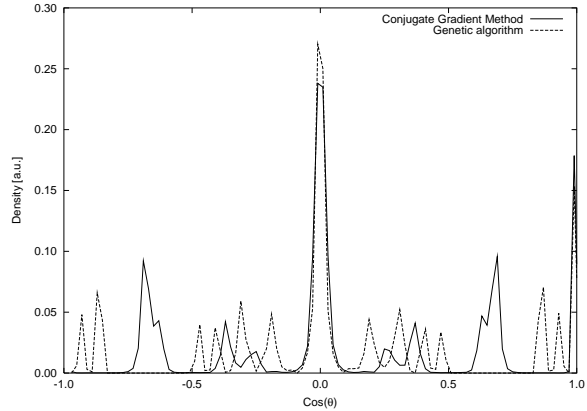, height=55mm}
\caption[]{Distribution of mutual orientations of every pair of molecules 
in the clusters. The genetic algorithm finds a unique structure (the peak 
at $cos(\theta) \approx 0.66$ is missing) if the number of the molecules 
in a cluster is greater than 80.}
\label{89-ori-compare}
\end{figure}

In the case of 89 and 137 $TeF_6$ clusters we have found a new local 
minimum on the PES which corresponds to quite a different structure with  
respect to the orientational order. Fig. \ref{89-ori-compare} represents 
these orientational structures produced with two different minimization 
algorithms, e.g. the  conjugate gradient and genetic algorithm.

\begin{figure}[htb]
\epsfig{file=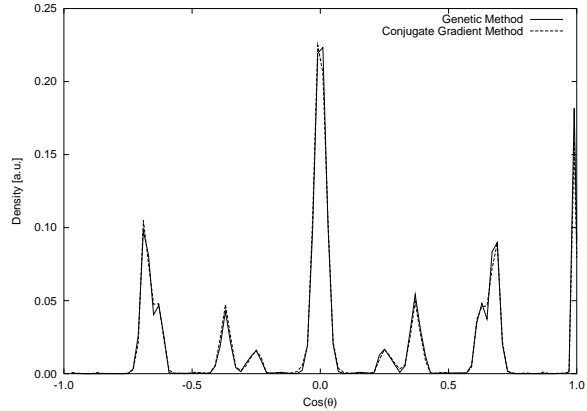, height=55mm}
\caption[]{The distributions of the mutual orientations of molecules show 
that smaller clusters (N $\leq$ 59) cannot pack in the new structure.}
\label{59-ori-compare}
\end{figure}

We have found that in the case of $TeF_6$ clusters with 27 and 59 
molecules such an orientation  distribution can not be found as it is 
seen in Fig. \ref{59-ori-compare}. This is a pronounced size effect. For 
the case of 89 $TeF_6$ clusters we perform a sequential molecular dynamic 
run to check if this new structure is stable. Starting from a low 
temperature $\approx$ 0.5 K, we increased the temperature up to 30 K 
and the structure is still stable.

\section{Conclusion}
Although the topography of the studied PES is very complicated, the newly 
developed algorithm has shown a great ability to find the "glocal" minimum. 
Mutations often boost this ability but in some cases it becomes worst. One 
can optimize the mutation parameters (percentage of mutated-cluster, 
mutation operator $M$, etc.) to obtain better results \cite{jordan}. 
Finally, we have found that the genetic algorithm can used to "clean" 
defects in structures optimized with techniques. For instance, the free 
molecular clusters have many surface molecules oriented improperly in 
comparison to the others even in the lowest energy configuration found 
with any other method. Such a procedure (cleaning) is very important if 
the density of states is needed. The general shortcoming of the method is 
its slowness, which makes its application limited.

\section*{Acknowledgments}
The authors thank for the partial finance support from the Scientific 
foundation of Plovdiv University.


\begin{thebibliography}{10}
\small
\bibitem{stoyan-01}
A. Proykova, S. Pisov, R. S. Berry,\\
J. Chem. Phys. {\bf 115} 8583 (2001).

\bibitem{Biswas} R. Biswas and D.R. Hamann\\
Phys. Rev. B {\bf 34}, 895 (1986).

\bibitem{stoyan-heronpress}
S. Pisov and A. Proykova,\\
Meetings in Physics {\bf 2}, 43 (2001) Herron Press.

\bibitem{Kunz-93}
R. E. Kunz and R. S. Berry, \\
Phys. Rev. Lett. {\bf 71}, 3987 (1993).

\bibitem{Doye-99}
J. P. K. Doye, N. A. Miller, D. J. Wales,\\
J. Chem. Phys. {\bf 110}, 6896 (1999).

\bibitem{Deaven} D.M. Deaven and K.M. Ho\\
Phys. Rev. Lett. {\bf 75}, 288 (1995).

\bibitem{Rama-02}
Jagtar S. Hunjan and R. Ramaswamy,\\
Int. J. Mol. Sci. 2002, {\bf 3}, 30.

\bibitem{Ani-99}
A. Proykova, R. Radev, Feng-Yin Li, R. S. Berry,\\
J. Chem. Phys. {\bf 110} 3887 (1999).



\bibitem{NR}
W.h. Press, B.P. Flannery, S.A. Teukolsky, W.T. Vetterling\\
{\it Numerical Recipes}, (Cambridge University Press, 1986).

\bibitem{jordan}
Jordan Yanev, \\
M. S. Thesis, University of Sofia 2001.

\end{thebibliography}
\end{document}